\documentclass[a4paper,12pt]{article} 
\usepackage[pdftex]{graphicx}
\usepackage[utf8]{inputenc}
\usepackage{euscript}
\usepackage{amsmath}
\usepackage{amsfonts}
\usepackage{amssymb}
\usepackage{wrapfig}

\newcommand{\eq}[1]{(\ref{eq:#1})}
\newcommand{\fig}[1]{Fig.\,\ref{fig:#1}}

\newcommand{\comment}[1]{}

\title{Fluxes and spectral indices of rare and abundant cosmic ray nuclei according to the NUCLEON space experiment}
\author{I.~A. Kudryashov\thanks{ilya.kudryashov.85@gmail.com}, A.~N. Turundaevskiy, D. E. Karmanov,\\
I. M. Kovalev, A.~A. Kurganov, A.~D. Panov, D.~M.  Podorozhny\\
\emph{Scobeltsyn Institute of Nuclear Physics},\\ \emph{Moscow State University}}
\date{}

\begin{document}
\maketitle
\sloppy

\begin{abstract}
In this paper the dependence of the spectra of cosmic ray nuclei on the charges of nuclei was studied, according to the data of the NUCLEON space experiment. First, we studied the dependence of the spectral index of magnetic rigidity spectra on the charge for abundant nuclei. Secondly, for the charge range $Z=9\div20$, the differences in the total spectra of rare odd and abundant even nuclei were studied. Using the GALPROP package, the inverse problem of CR propagation from a source (near supernova) to an observer was solved, a component-by-component spectrum in the source was reconstructed, and it was shown that a systematic change in the spectral index in the source exist. It is supposed that this change may be interpreted as incomplete ionization of cosmic rays at the stage of acceleration in the supernova remnant shock. The ratio of the total spectra of magnetic rigidity for low-abundance odd and abundant even nuclei from the charge range $Z=9\div20$ is obtained, and it was shown that the spectra of odd rare nuclei are harder than the stpectra of abundat even nuclei in the rigidity range 300--10000~GV.
\end{abstract}

\section{Introduction}

The study of the chemical composition and energy spectrum of galactic cosmic rayss (GCR) provides important information about the physics of galactic sources of cosmic rays -- supernovae remnants. The GCR spectra observed near the earth differ significantly from the spectra in the sources, since the transport of galactic cosmic rays with a magbetic rigidity of less than 100~TV has a diffusion character, with a diffusion coefficient that depends on the magnetic rigidity. In this case, possible differences between the spectra of abundant and rare nuclei are a separate important issue since the abundant nuclei are mainly primary character but rare odd nuclei are supposed to have essential secondary part. To obtain the spectra in the sources, it is necessary to solve the inverse problem of CR propagation through the galactic interstellar medium.

\section{Solution of the inverse propagation problem}

To solve the inverse propagation problem, the results of modeling with the well-known GALPROP package were used to solve the direct propagation problem \cite{C1}. The inverse propagation problem is reduced to solving a sequence of direct problems. To solve the direct problems a set of parameters of the diffusion propagation model was used, obtained by the HELMOD collaboration team based on Monte Carlo simulation \cite{C2}.

Due to the fact that the characteristic quantity for the mechanisms of CR propagation is the magnetic rigidity $R$, but in space experiments based on ionization calorimeters the energy per particle is measured, the energy range of measurements of only one experiment is insufficient for constructing a well statistically valid model. Therefore, in order to expand the range of experimental data, two experiments are included in the analysis -- NUCLEON \cite{C3,C4,C5} (two measurement methods -- kinematic method KLEM and classical ionization calorimeter method) and ATIC \cite{C6} (ionization calorimeter). A joint analysis of data from different experiments was previously used in \cite{C7}.

To solve the problem of transition from the spectrum observed in the local observations to the spectrum in the source, the following technique was used.
The GALPROP package calculates particle transport in the isotropic diffusion approximation, where the diffusion coefficient $D$ depends on the magnetic hardness as follows
\begin{equation}
 D = D_0\times(R/R_0)^\delta
 \label{eq:Diffusion}
\end{equation}
For each $D_0$ and $\delta$ from Eq. \eq{Diffusion}, in the range of three standard deviations from the value obtained by the Monte Carlo method, a family of observed spectra was built with different spectrum slope index at the source ($\gamma_{star}$). The experimental data was approximated by a power law, where the spectral index $\gamma_{star}$ is a free parameter. For each such family, the dependence of the spectral index of the spectrum in the source on the slope of the observed spectrum was obtained, the resulting dependence was approximated by a straight line, and one can obtain the spectrum in the source from the observed slope of the spectrum.

\begin{figure}
\includegraphics[width=\textwidth]{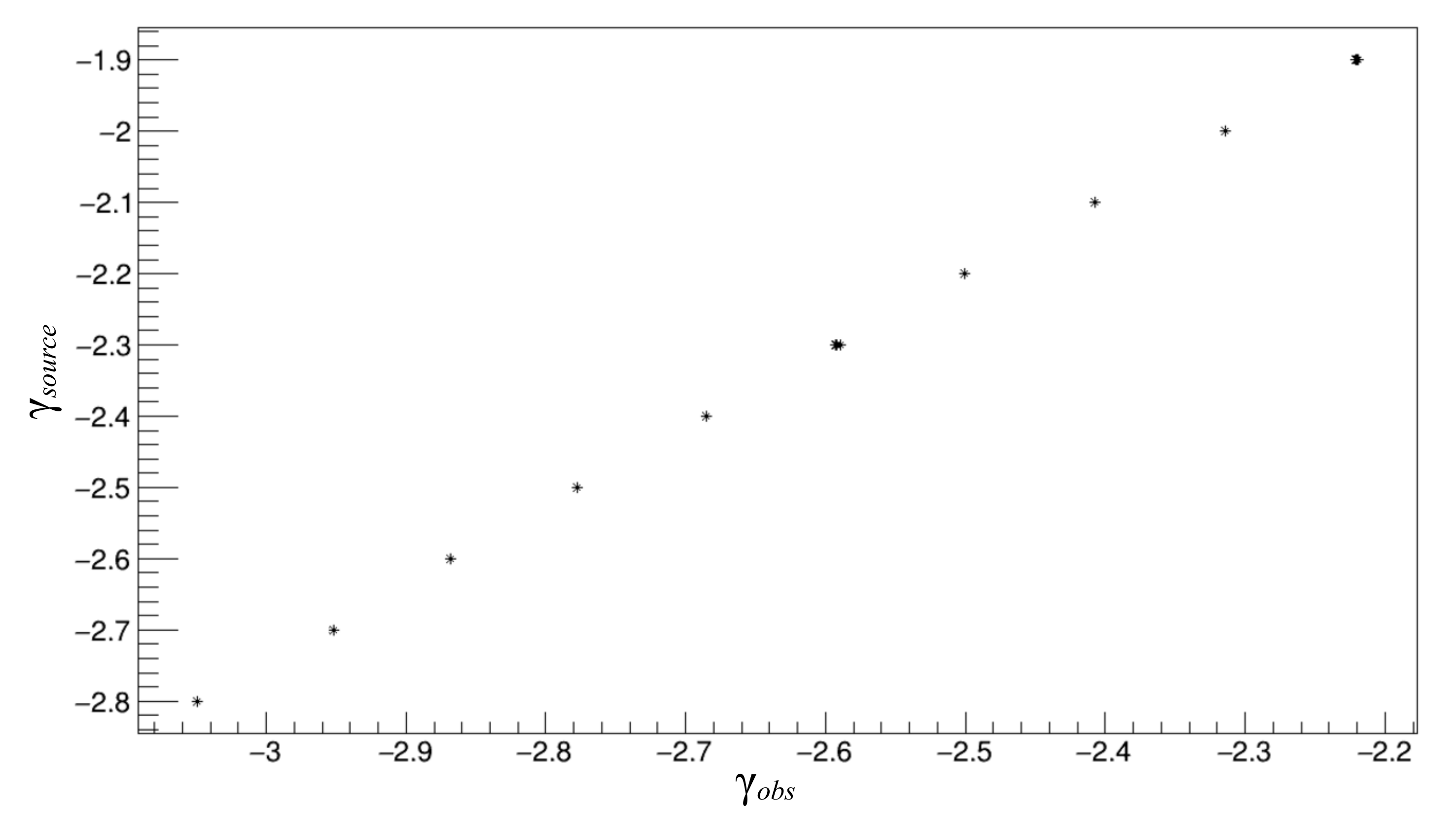}
 \caption{Dependence of the spectral index in the source on the observed spectral index for fixed values of $D_0$ and $\delta$.}
 \label{fig:Fig1}
\end{figure}

The dependence of the spectral index in the source on the observed spectral index is shown in \fig{Fig1} for some fixed model parameters. Such modeling was carried out for all abundant nuclei and the spectral indeces of the CR spectra in the source were obtained. The obtained values are very sensitive to the parameters of the CR propagation model, primarily $D_0$ and $\delta$. The dependence on the parameter $\delta$ is expected and predicted by the simplified \emph{Leaky box} propagation model \cite{C8}, while the dependence on $D_0$ can be observed only in the more sophisticated diffusion propagation model, but the \emph{Leaky box} model is not sensitive to it.

\begin{figure}
\includegraphics[width=\textwidth]{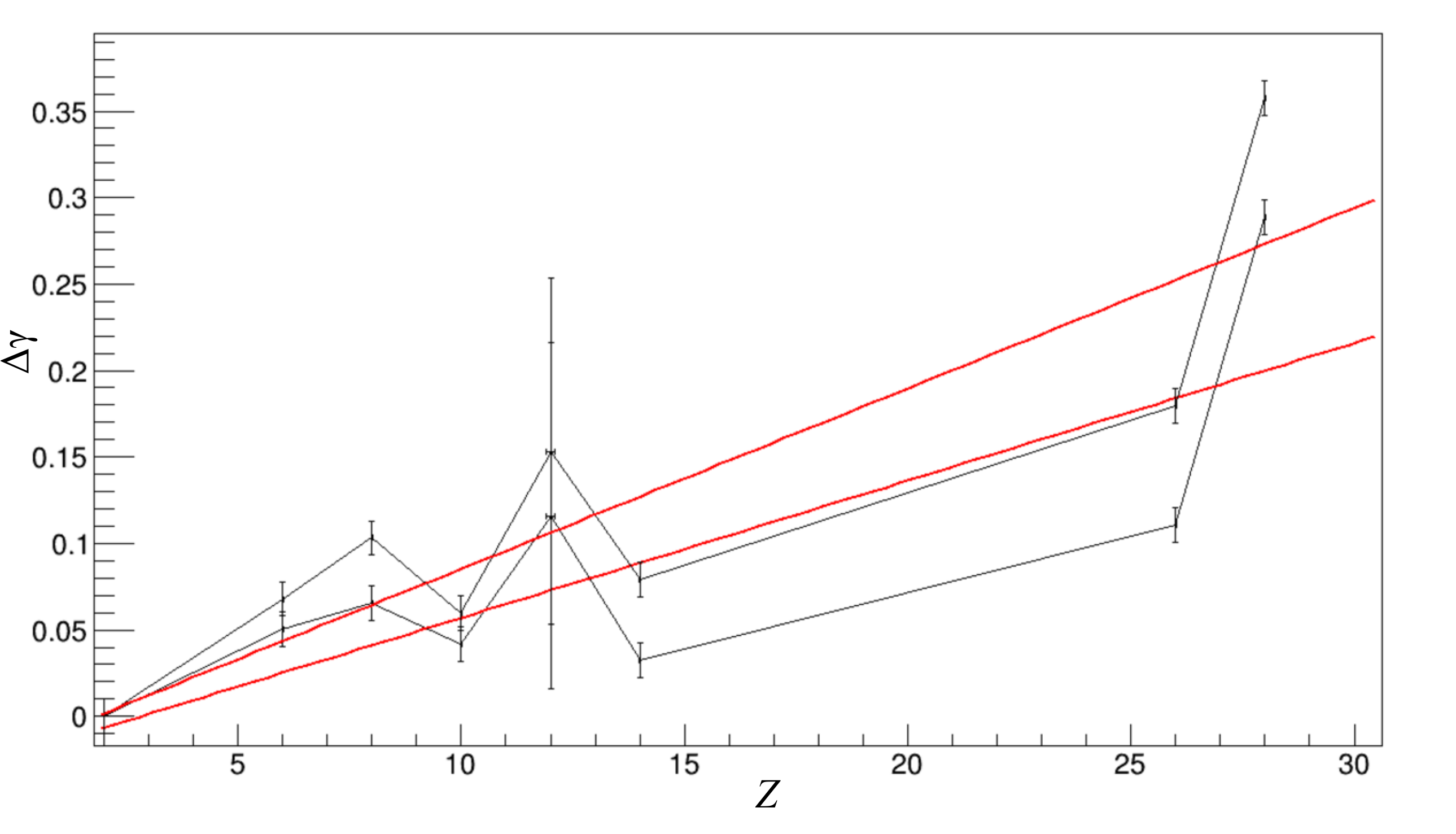}
 \caption{Dependence of $\Delta\gamma$ on the nuclear charge.}
 \label{fig:Fig2}
\end{figure}

To reduce the systematic uncertainty, the difference between the spectral indices of the spectra of helium and other abundant elements in the measured range of $D_0$ and $\delta$ was plotted. \fig{Fig2} shows the difference between the spectral index of helium and the rest of the abundant primary nuclei, including nickel. Statistical errors are recalculated from the errors of the measured spectral index of the spectra, the upper and lower curves correspond to the relations of $D_0$ and $\delta$, at which the calculated difference is minimal and maximal, respectively. They correspond to two linear approximations, the flatter one has a slope $\Delta\gamma/Z = (6.9 \pm 0.7)\times 10^{-3}$. Thus, the direct dependence of the spectral index in the source on the charge of the cosmic ray nucleus in the range from 25~GV to 1500~GV is statistically significant at least at the level of 9 standard deviations. This result confirms the earlier observation of the ATIC collaboration \cite{C6} with higher statistical significance. An indication of this effect was also obtained from the AMS-02 and HEAO-3-C2 data \cite{C9} after taking into account of solar modulation in the HelMod package.

This result can be interpreted in several ways:
\begin{itemize}
 \item
    as a feature of the medium around the supernova in which the acceleration of cosmic rays takes place. Such a medium could be enriched in heavy elements from the stellar wind before the supernova explosion. The effect is then due to the radial layering of various elements \cite{C10,C11};
 \item
    as incomplete ionization of heavy cosmic ray nuclei during acceleration on shock waves during supernova explosions;
 \item
    as a dependence of the injection efficiency during acceleration on the A/Z ratio \cite{C12}.
\end{itemize}

This result, obtained with high statistical significance for all nuclei of classical nucleosynthesis, will allow us to significantly refine the mechanisms of supernova explosions and subsequent CR acceleration on shock waves generated by these explosions.

\begin{figure}
\begin{center}
\includegraphics[width=0.7\textwidth]{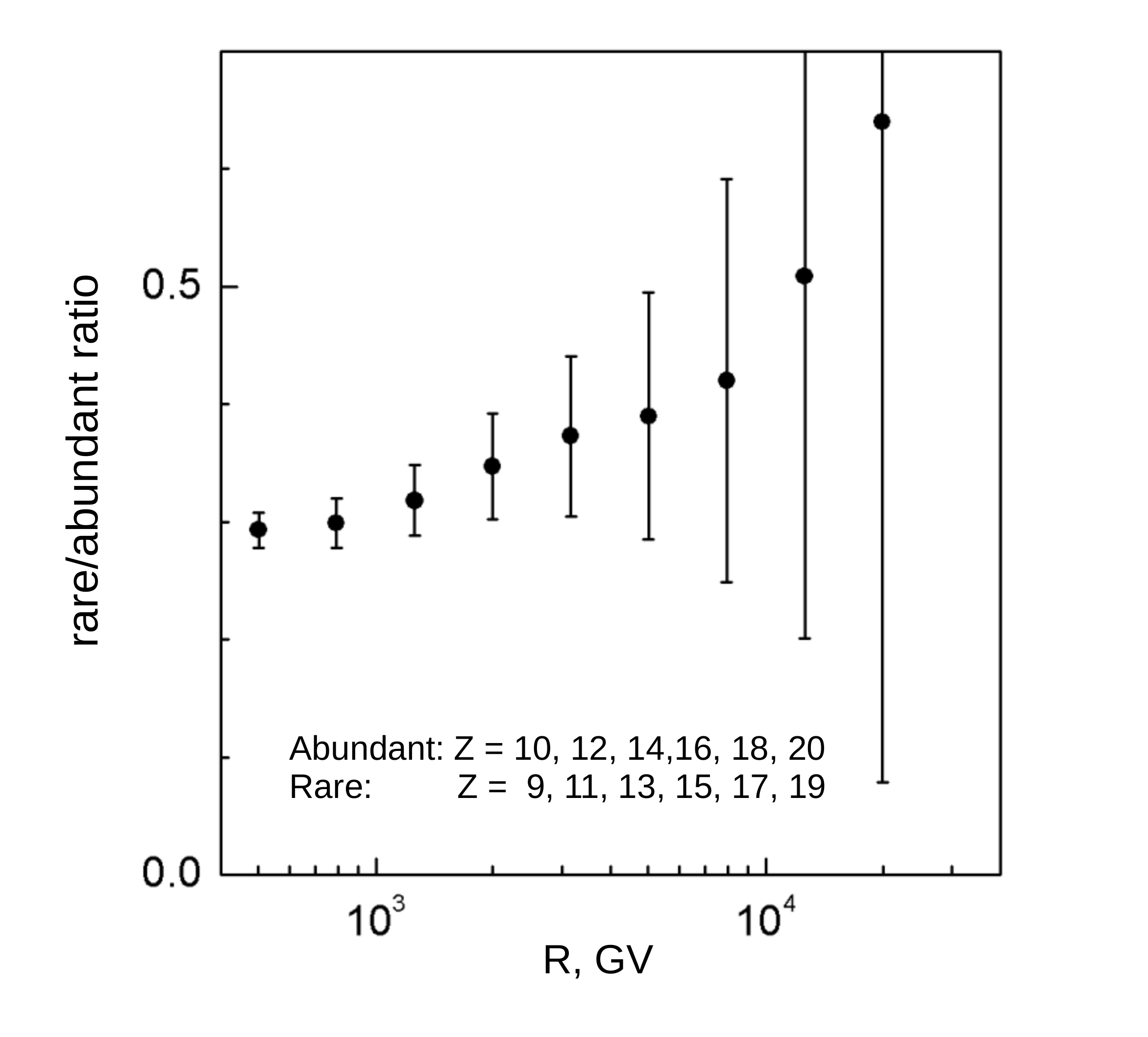}
\end{center}
 \caption{Ratio of spectrum of rare to spectrum of abundant nuclei.}
 \label{fig:Fig3}
\end{figure}

\section{Spectra of rare odd nuclei and even abundant nuclei}

\begin{figure}
\begin{center}
\includegraphics[width=\textwidth]{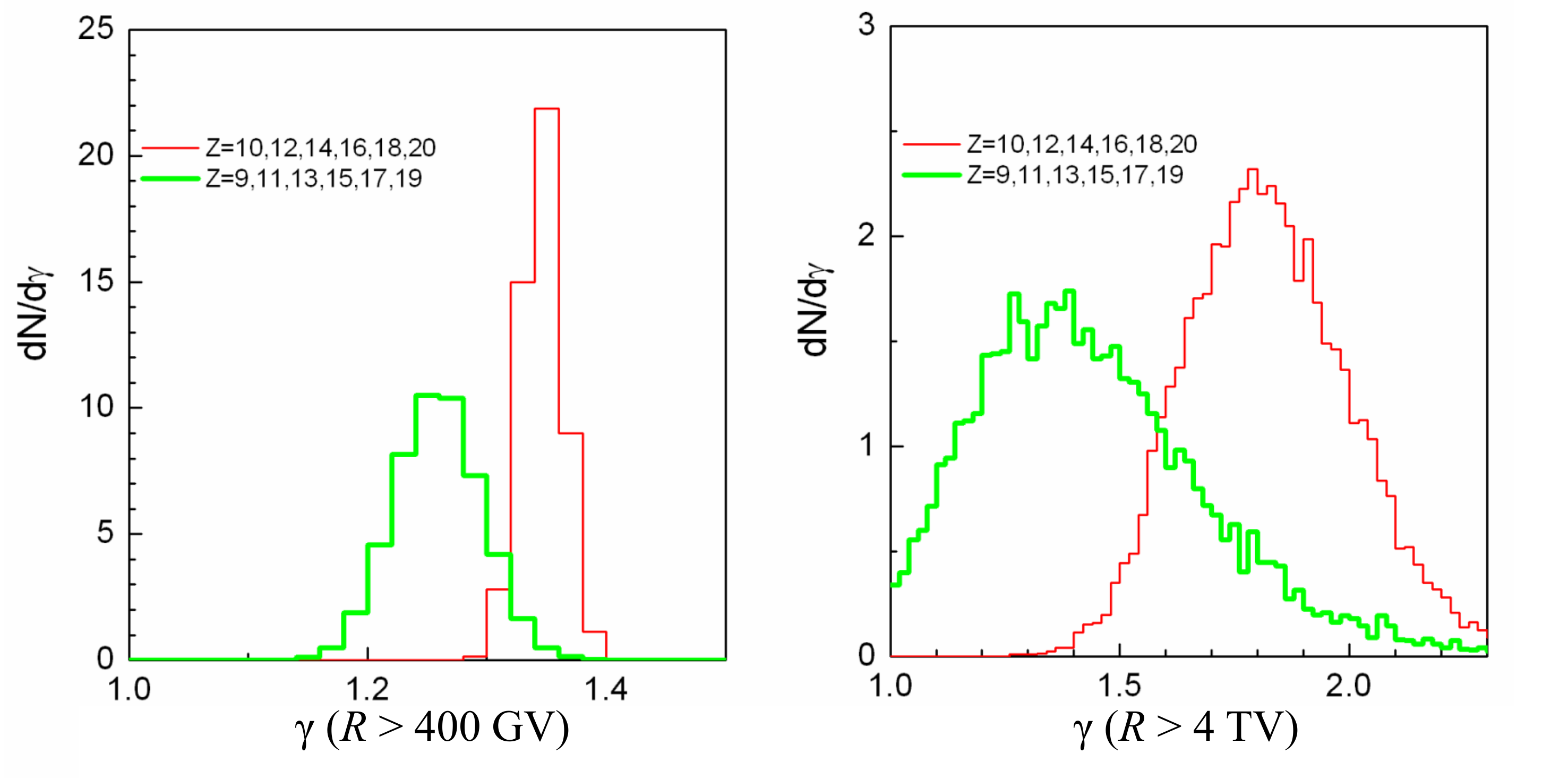}
\end{center}
 \caption{Distributions of the reconstructed spectral indeces of the integral spectrum based on the results of Monte Carlo modeling of the distribution of errors for thresholds of 400~GV and 4~TV}
 \label{fig:Fig4}
\end{figure}

When analyzing the information obtained in the NUCLEON experiment, the question arises whether the spectra of rare and abundant nuclei differ. Nuclei with an even charge, starting with carbon, are abundant, and odd nuclei (nitrogen, fluorine, etc.) are rare. It is natural to look for possible differences in the spectra in the magnetic rigidity scale. The complexity of working with the spectra of rare nuclei with the data of NUCLEON experiment lies in the fact that, due to the finite charge resolution, the intensity transfer from neighboring abundant nuclei to the flux of a rare nucleus is possible. Therefore, it was necessary to use more strict than usual criteria for identifying nuclei: only nuclei with $Z = Z_0 \pm 0.4$ were selected both for odd and even nuclei. To increase the statistics, the total fluxes of rare nuclei with charges 9, 11, 13, 15, 17, 19 and abundant nuclei with charges 10, 12, 14, 16, 18, 20 were compared. The resulting ratio of the spectra is shown in \fig{Fig3}. This ratio increases with increasing magnetic rigidity, which indicates a harder spectrum of rare nuclei. However, the statistical errors are large and a more detailed analysis is required to determine the statistical significance of this growth. The maximum likelihood method was used at different thresholds for magnetic rigidity, to determine the spectral index for both rare and abundant nuclei. \fig{Fig4} shows the distributions of the absolute value of the integral spectral index for thresholds of 400~GV and 4~TV. It turned out that the spectral indices of abundant and rare nuclei at the magnetic rigidity threshold of 400~GV are $2.34 \pm 0.015$ and $2.24 \pm 0.025$, respectively. Thus, there is a difference in the spectral indices at the level of 3.4$\sigma$.

\section{Conclusion}

The performed analysis of the experimental data of the NUCLEON experiment demonstrated the presence of hitherto unknown effects in the spectra of cosmic rays. The direct dependence of the spectral index in the source on the charge of the cosmic ray nuclei, the existence of which was previously indicated in the ATIK experiment, has been confirmed with high statistical significance. The effect was confirmed in the range from 25 GV to 1500 GV, and the spectra of cosmic ray nuclei in the source were estimated using the GALPROP package. The difference between the spectra of rare odd and abundant even nuclei in the range of charges from 9 to 20 is demonstrated. The spectrum of rare odd nuclei is more rigid, which contradicts to naive expectations based on the fact, that the fraction of the secondary component among rare odd nuclei is relatively high.


\begin{thebibliography}{}
 \bibitem{C1}
    https://galprop.stanford.edu/

 \bibitem{C2}
    https://www.helmod.org/index.php

 \bibitem{C3}
    Turundaevskiy A.N., Vasiliev O.A., Karmanov D.E. et al. // Bull. Russ. Acad. Sci. Phys. 2021. V. 85. No. 4. P. 353.

 \bibitem{C4}
    Atkin E.V., Bulatov V.L., Vasiliev O.A. et al. // Bull. Russ. Acad. Sci. Phys. 2019. V. 83. No. 8. P. 977.

 \bibitem{C5}
    Panov A.D., Atkin E.V., Bulatov V.L.et al. // Bull. Russ. Acad. Sci. Phys. 2019. V. 83. No. 8. P. 980.

 \bibitem{C6}
    Panov A.D., Zatsepin V.I., Sokolskaya N.V. // Bull. Russ. Acad. Sci. Phys. 2015. V. 79. No. 3. P. 285.

 \bibitem{C7}
    Kudryashov I.A., Kovalev I.M., Kurganov A.A. et al. // Bull. Russ. Acad. Sci. Phys. 2021. V. 85. No. 4. P. 379.

 \bibitem{C8}
    Ptuskin V., Strelnikova O., Sveshnikova L. // Proc. 31st ICRC (Łodz, 2009). P. 593.

 \bibitem{C9}
    Boschini M.J., Della Torre S., Gervasi M. et al. // Astrophys. J. 2018.  V.858. P.61.

 \bibitem{C10}
    Ohira Y. and Ioka K. // Astrophys. J. Lett. 2011. V.729. P.L13.

 \bibitem{C11}
    Ohira Y., Kawanaka N., Ioka K. //  Phys. Rev. D, 2016. V. 93. 083001

 \bibitem{C12}
    Malkov M.A. // Phys.Rev. E. 1998. V. 58, 4911
\end{thebibliography}
\end{document}